\documentclass[fleqn,usenatbib]{mnras}

\usepackage{newtxtext,newtxmath}

\usepackage[T1]{fontenc}

\DeclareRobustCommand{\VAN}[3]{#2}
\let\VANthebibliography\thebibliography
\def\thebibliography{\DeclareRobustCommand{\VAN}[3]{##3}\VANthebibliography}

\usepackage{graphicx}	
\usepackage{amsmath}	
\usepackage{array}
\usepackage{booktabs}
\usepackage{pdflscape}

\title[FAST Observations: A Scarcity of Large PAHs?]{The Missing Giant: Do FAST Spectroscopic Observations Reveal a Scarcity of Large Polycyclic Aromatic Hydrocarbons in Astronomical Environments?}

\author[Shao et al.]{
Yi Shao,$^{1}$
Yong Zhang,$^{1,2}$\thanks{E-mail: zhangyong5@mail.sysu.edu.cn (YZ)}
Xu-Jia Ouyang,$^{3,1}$
Chuan-Peng Zhang$^{4}$
\\
$^{1}$School of Physics and Astronomy, Sun Yat-sen University, 2 Daxue Road, Tangjia, Zhuhai, Guangdong Province, China\\
$^{2}$CSST Science Center for the Guangdong-Hongkong-Macau Greater Bay Area, Sun Yat-Sen University, Guangdong Province, China\\
$^{3}$College of Physics, Guizhou University, Guiyang 550025, China\\
$^{4}$National Astronomical Observatories, Chinese Academy of Sciences, Beijing 100101, China
}

\date{Accepted XXX. Received YYY; in original form ZZZ}

\pubyear{\the\year{}}

\begin{document}
\label{firstpage}
\pagerange{\pageref{firstpage}--\pageref{lastpage}}
\maketitle

\begin{abstract}
The search for large polycyclic aromatic hydrocarbons (PAHs) with over 100 carbon atoms is crucial to resolving the origin of unidentified infrared emission (UIE) bands. These bands are commonly observed in nebulae and the interstellar medium, yet their spectroscopic assignment has remained unknown for decades. Using the Five-hundred-meter Aperture Spherical Radio Telescope (FAST), the world’s most sensitive instrument operating in the decimeter-wavelength range, we conducted a search for rotational transitions of large, quasi-symmetric PAHs. Our sample included two prototypical UIE sources, NGC 7027 and TMC-1, along with a non-UIE source, IRC+10216, for comparison. A matched filter technique was employed to isolate comb-like spectral features from quasi-symmetric PAHs containing 138 to 194 carbon atoms in the FAST spectra. This method significantly enhanced detection sensitivity to these astrophysically critical molecular signatures.
 Although no such features were detected, we derived upper limits on the abundance of large PAHs based on simplifying assumptions. These upper limits are lower than the values predicted by theoretical models, which might tentatively suggest that large PAHs may not be the primary carriers of UIE bands.  However, this conclusion should be treated as tentative, given that it rests on simplistic assumptions which have not been empirically validated.
\end{abstract}




\section{Introduction}
The Unidentified Infrared Emission (UIE) bands at 3.3, 6.2, 7.7--7.9, 8.6, 11.3, and 12.7 $\mu$m have been detected in numerous astronomical objects \citep[e.g.,][]{2014IAUS..297..187P}. These characteristic features are observed across diverse astrophysical environments, ranging from galactic nuclei to the diffuse interstellar medium (ISM), and from young stellar objects to evolved stellar systems such as protoplanetary nebulae and planetary nebulae \citep[e.g.,][]{2013IAUS..292..288K,2020NatAs...4..339L}. The carriers responsible for these emission bands are hypothesized to represent a significant component of interstellar dust, potentially holding critical information about the interstellar carbon cycle \citep[see][for a comprehensive reivew]{2008ARA&A..46..289T}. Their ubiquitous presence suggests a fundamental role in cosmic carbon processing mechanisms throughout the universe.

The Unidentified Infrared Emission (UIE) bands are widely attributed to C-H and C–C vibrational modes in aromatic compounds \citep{1981MNRAS.196..269D},  designated as Aromatic Infrared Bands (AIBs). Numerous hydrocarbon materials have been proposed as potential carriers, including hydrogenated amorphous carbon \citep{1983MNRAS.205P..67D}, quenched carbonaceous composites \citep{1987ApJ...320L..63S}, polycyclic aromatic hydrocarbon (PAH) molecules \citep{1984A&A...137L...5L,1985ApJ...290L..25A,1989ARA&A..27..161P},
soot and carbon
nanoparticles \citep{2008ApJ...677L.153H},
coal-like substances \citep{1989ESASP.290..207P}, and petroleum fractions \citep{2003IJAsB...2...41C,2013MNRAS.429.3025C}.
These UIE bands typically appear superimposed on broad plateau features spanning 6--9 $\mu$m and 10--15 $\mu$m, attributed to blended emission from bending modes in mixed alkyl/alkenyl groups \citep*{2001ApJ...554L..87K}, indicating significant aliphatic contributions. This spectral complexity has motivated the  Mixed Aromatic/Aliphatic Organic Nanoparticle (MAON) model \citep{2011Natur.479...80K,2013ApJ...771....5K}, which exhibits fundamentally distinct molecular architectures from planar PAHs through disordered structures where aromatic rings ($sp^{2}$-bonded) interconnect via aliphatic chains ($sp^{3}$-bonded). It has been demonstrated that the petroleum and coal model of the UIE demonstrates strong convergence and close integration with the MAON model \citep{2020Ap&SS.365...81C}.




Among these proposed UIE carriers, the PAH hypothesis has emerged as the dominant theoretical framework \citep*{2008ARA&A..46..289T}, with AIBs frequently termed ``PAH features'' in astronomical literature. However, infrared spectra alone cannot validate the PAH hypothesis \citep{2015ApJ...798...37Z}. Complementary ultraviolet and optical spectroscopic searches targeting PAH electronic transitions have established upper limits on specific PAH abundances, but yielded no definitive detections \citep{2011MNRAS.412.1259G,2011A&A...530A..26G}.

Rotational spectroscopy in the radio band provides definitive molecular fingerprints for identifying chemical species in astronomical environments.
However, it is challenging to observe the rotational transitions of PAHs. Most neutral PAHs exhibit relatively low (or zero) permanent dipole moments, resulting in intrinsically weak rotational transitions. Moreover, astronomical PAHs are predominantly large molecules lacking symmetry, leading to complex rotational emission patterns fragmented into numerous weak spectral lines \citep{2008ARA&A..46..289T}.
An exception is corannulene (C$_{20}$H$_{10}$), a bowl-shaped PAH featuring a pentagonal carbon ring surrounded by five hexagons. The molecular curvature endows C$_{20}$H$_{10}$ with a permanent dipole moment of 2.1 D along its symmetry axis, making it a promising candidate for astronomical detection via rotational spectroscopy. 
However, spectral observations have not resulted in definite detections of C$_{20}$H$_{10}$ \citep{2005JAChS.127.4345L,2009MNRAS.397.1053P}.

The search for PAH rotational lines has advanced significantly in recent years. 
Theoretical models suggest that highly symmetric ``grand PAHs'' may dominate interstellar populations due to enhanced ISM survivability, potentially leading to their overabundance \citep{2013RvMP...85.1021T}. Nitrogen-substituted PAHs, which are strongly polar and potentially prevalent in the ISM \citep{2005ApJ...632..316H}, may retain quasi-symmetric structures, producing distinctive comb-like rotational spectra with regular line spacing. This spectral simplicity reduces signal dilution and enables matched-filtering techniques \citep{2014MNRAS.437.2728A}, significantly boosting detection sensitivity. The predictable line spacing further facilitates systematic blind searches for unknown molecules by scanning spacing parameters, eliminating the need for exhaustive quantum-chemical modeling or laboratory rotational constants.
Stacking and matched filter analysis of sensitive molecular cloud spectra have detected specific 2--7 ring PAHs \citep{2021Sci...371.1265M,2021NatAs...5..181B,2021A&A...649L..15C,2024Sci...386..810W,2025NatAs...9..262W,2025ApJ...984L..36W}. Applying this technique, \citet{2025MNRAS.538.2084M} report stringent abundance upper limits for bowl-shaped PAHs C$_{20}$H$_{10}$ and sumanene (C$_{21}$H$_{10}$).

Nevertheless, due to their inherently lower thermal stability relative to larger PAHs, the small PAHs discussed in these studies fall outside the classification of ``grand PAHs''. Critically, their characteristic infrared vibrational spectra exhibit significant discrepancies when compared to the 
UIE bands. According to the prevailing PAH hypothesis, the carriers responsible for the UIE features are postulated to consist of much larger PAH molecules, typically comprising hundreds of carbon atoms \citep[e.g.,][]{2016A&A...590A..26C, 2021ApJ...918....8K}.

In this study, we conduct a search for rotational transitions of large PAHs 
(containing $>130$ carbon atoms)
toward UIE sources using the Five-hundred-meter Aperture Spherical radio Telescope (FAST), which offers unparalleled sensitivity in the decimeter-wavelength range. 
We employ the matched-filter analysis method described by \citet{2015MNRAS.447..315A}. The targeted sources include NGC 7027, TMC-1, and IRC+10216. The first two are established UIE emitters, while the latter lacks UIE bands and serves as a non-UIE comparison source.
The paper is structured as follows. Section 2 presents the method for searching PAH rotational transitions, with accuracy confirmed through controlled experiments. Section 3 details the observational parameters and data processing workflow. Section 4 reports blind signal search results for  three astrophysical sources. Section 5 establishes upper abundance limits based on these results and compares them with theoretical predictions. Section 6 summarizes the findings.

\section{Methodology}

An ideal planar rigid symmetric top emits a comb-like rotational spectrum characterized by equally spaced lines with frequency spacing
\(\Delta \nu_{\text{comb}} = \hbar / (2\pi I_3)\), where \(I_3\) is the largest moment of inertia, which is precisely twice as large as the other two degenerate moments.
Using computed rotational constants for coronene and circumcoronene \citep{2005ApJ...632..316H}, 
\citet{2014MNRAS.437.2728A} derive the following scaling relation for the principal moment of inertia of disc-like PAHs:
\begin{equation}
I_3 \approx 1.5 \times 10^4 \left( \frac{N_{\rm C}}{54} \right)^2 \, \text{amu} \, \text{{\AA}}^2,
    \label{eq:1}
\end{equation}
where \(N_{\rm C}\) represents the carbon atom count.
Larger molecules exhibit a greater moment of inertia, leading to rotational spectral peaks at lower frequencies. 
Equation~\ref{eq:1} is consistent with the empirical relation between the PAH radius and 
$N_{\rm C}$ as 
presented in the textbook by \citet{Ttextbook}.
Considering the frequency range covered by our FAST observations (1.6--3.2 GHz), the current search is focused on PAHs with $N_{\rm C}>130$.

To demonstrate the methodology of searching for comb-like spectra, we simulated the rotational spectra of PAHs with 160 carbon atoms, where the comb spacing was configured as 6.6 MHz. A probability distribution framework following \citet{2011MNRAS.411.2750S} was implemented,
\begin{equation}
     P(J) \propto J^2 \exp\left[-\left(\frac{J}{200}\right)^2\right],
\end{equation}
where $J$ is an integer and the corresponding emission
frequency follows $\nu_J = (J + 1/2)\Delta \nu_{\text{comb}}$.
The probability distribution function $P(J)$ delineates the peak emission characteristics of PAH rotational spectral lines. At the rotational quantum number $J$ corresponding to the maximum probability, the associated emission frequency represents the peak frequency of rotational line emission. The probability distribution function analysis reveals that the rotational quantum number $J$ exhibits a pronounced peak at $J = 147$, so the spectrum peaks near 2 GHz.

 \begin{figure}
     \centering
     \includegraphics[width=0.99\linewidth]{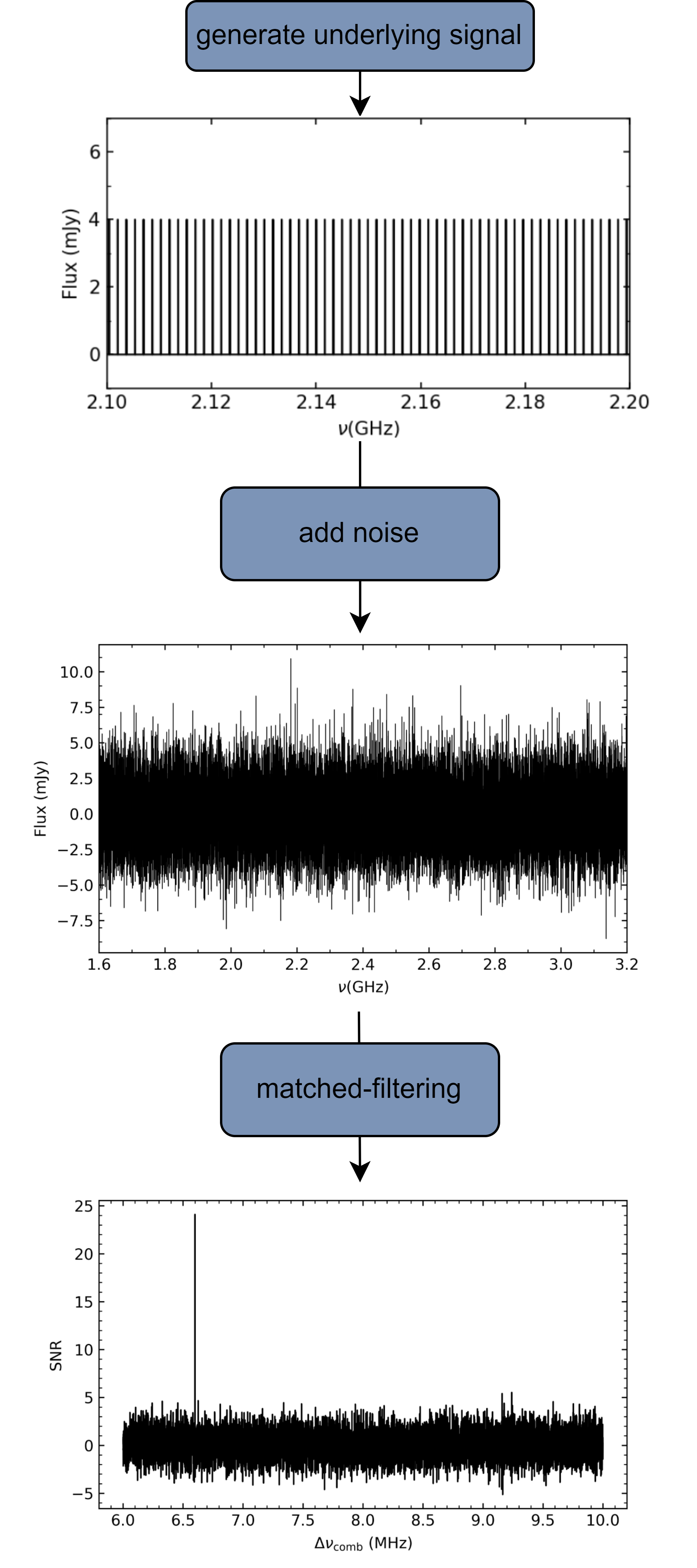}
     \caption{Flowchart outlining the methodology for searching for comb-like spectral features of PAHs.
The simulated spectrum (middle panel) is generated by adding noise to the PAH spectrum (upper panel). The SNR as a function of $\Delta\nu_{\text{comb}}$ (lower panel) 
is obtained through the matched-filtering technique.
    }
     \label{fig:flow chart}
 \end{figure}

Figure~\ref{fig:flow chart} depicts a flowchart outlining our methodological framework. 
To simulate the observed spectra, we first generate a comb-like spectrum spanning a 1.6 GHz bandwidth, where each spectral line is broadened using a 2-MHz-wide Gaussian function. Subsequently, Gaussian random noise is added with a root-mean-square (rms) amplitude of approximately one-third the mean line intensity. These assumptions are employed to simulate the spectra and demonstrate the methodology, and do not introduce uncertainty into realistic spectral searches. While windows of varying widths or profiles could be employed, the resulting outcomes would remain comparable.
Under this configuration, individual spectral lines remain undetectable at the $3\sigma$ confidence threshold
(as illustrated in the middle panel of Figure~\ref{fig:flow chart}).

The blind search algorithm \citep[see][for detailed methodology]{2015MNRAS.447..315A} scans over possible comb tooth spacings $\Delta\nu_{\text{comb}}$, with the resultant Signal-to-Noise Ratio (SNR) spectrum is presented in the bottom panel of Figure~\ref{fig:flow chart}. For a comb-like spectrum with tooth spacing $\Delta\nu_{\text{comb}}$, the SNR is calculated by summing the intensities at the expected comb frequencies, then normalizing by the product of the rms fluctuation and the square root of the number of spectral lines within the bandwidth. This analysis demonstrates robust reconstruction of the input comb structure. In this representative case, an SNR of $\approx 20$ is achieved, while the known line spacing is precisely recovered. 
Notably, the spectral pattern is characterized by a single  parameter---the line spacing $\Delta\nu_{\text{comb}}$, which directly correlates with the carrier molecule's physical dimensions. This parametric simplicity facilitates efficient blind searches for PAHs of varying sizes, as opposed to targeting specific molecular species.


\begin{figure}
	\includegraphics[width=\columnwidth]{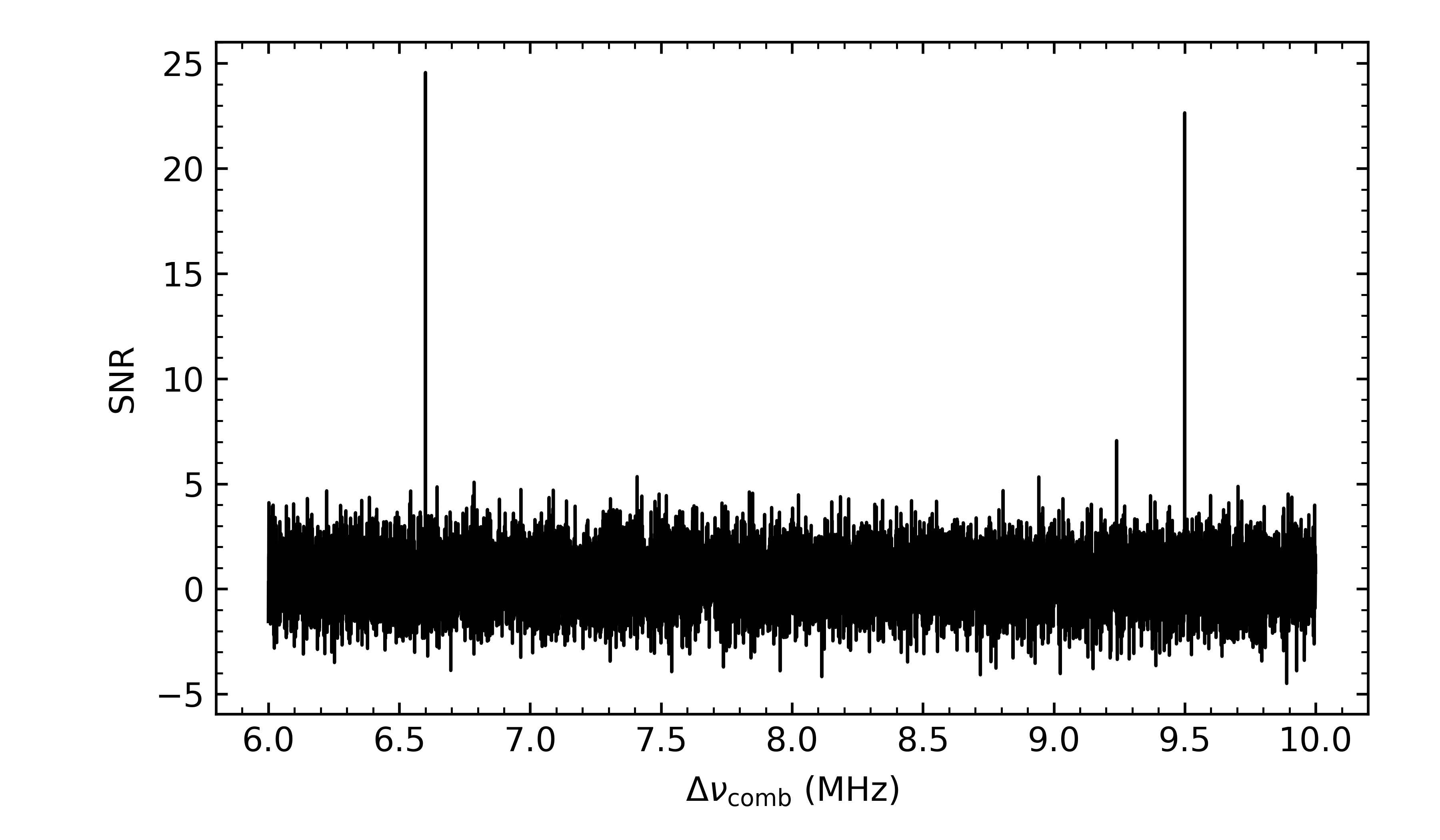}
    \caption{SNR as a function of $\Delta\nu_{\text{comb}}$
    obtained through the matched-filtering technique.
    The input spectrum is identical to that in Fig.~\ref{fig:flow chart}, but is superimposed with a comb-like spectrum characterized by a tooth spacing of 9.5 MHz.}
    \label{fig:SNR for two comb}
\end{figure}
This analytical methodology enables the simultaneous detection of multiple PAHs exhibiting distinct rotational constants. Figure~\ref{fig:SNR for two comb} presents simulation results demonstrating this capability, extending the basic case shown in Fig.~\ref{fig:flow chart} through the introduction of an additional harmonic comb structure characterized by 9.5 MHz tooth spacing.
The results confirm accurate spectral reconstruction with two clearly resolved peaks. This analytical approach provides unambiguous molecular discrimination based on unique rotational signatures, enabling population studies of PAH molecules in astrophysical environments through their characteristic radio spectral fingerprints.

\section{OBSERVATIONS AND DATA REDUCTION}

\subsection{Observed Target Sample}

NGC 7027 stands as one of the earliest sources where the UIE was detected \citep*{1977ApJ...217L.149R}. Its elevated ionization state and dense dust environment render it a pivotal target for astrochemical investigations. Within such circumstellar environments, the multilayered dust configuration typically provides photoprotective shielding against ultraviolet radiation, thereby enhancing the survival probabilities of molecules and facilitating the formation and preservation of large molecules.
With the UIE bands attributed to PAHs, the average $N_c$ in NGC 7027 is estimated to be $149 \pm 34$ \citep{2020MNRAS.494..642M}, positioning this source as an ideal candidate for searching large PAHs.
 
TMC-1, a dense and cold molecular core within the nearby star-forming Taurus Molecular Cloud complex, is recognized as one of the most chemically rich interstellar environments. Characterized by extremely low temperatures and high densities, conditions favorable for the formation of complex organic molecules, small PAHs with $N_{\rm C}<30$ have been detected in this source via their rotational spectra~\citep{2021Sci...371.1265M,2021NatAs...5..181B,2021A&A...649L..15C,2022ApJ...938L..12S,2024Sci...386..810W,2025ApJ...984L..36W}, justifying the search for large PAHs here.
 
IRC+10216 is a carbon-rich asymptotic giant branch star. Although a number of molecules have been detected toward this source, the UIE bands are conspicuously absent \citep[e.g.,][]{2023JApA...44...47A}. Therefore,
IRC+10216 is likely a non-PAH source.
Observations of IRC+10216 were conducted to function as a comparative reference source. 
If PAH rotational emission were to be detected in the spectra of NGC 7027 and TMC-1, it should not manifest in that of IRC+10216.

\subsection{Observations}

The observations were conducted on 2022 November 18, utilizing the newly installed cryogenic ultrawideband (UWB) receiver of the FAST \citep{2022RAA....22k5016L,2023RAA....23g5016Z}. An ON-OFF position switching mode was adopted with a total integration time of 1 hour. The UWB receiver operates within the frequency range of 500--3300 MHz
and is divided into four sub-bands, each consisting of 1,048,576 channels, yielding a frequency resolution of 1049.04 Hz. Within the 500--3300 MHz range, the antenna gain varies from approximately 
14.3 to 7.7 K Jy\(^{-1}\).
The aperture efficiency ranges from about 0.56 to 0.30, while the system temperature spans between 88 and 130 K.
 The half-power beamwidth ranges approximately from \(7.6^\prime\) to \(1.6^\prime\). For our analysis, we focused exclusively on data within the  1.6 to 3.2 GHz frequency range. These represent the most sensitive spectra of the three sources in this frequency range to date.

\subsection{Data Reduction}

Flux calibration was performed using noise diode signals. The antenna temperature \( T_A \) was derived from the measured power during the "Cal on" (\( P_{\text{calon}} \)) and "Cal off" (\( P_{\text{caloff}} \)) phases, combined with the noise diode temperature (\( T_{\text{cal}} \)):
\begin{equation}
T_A = w_1 \cdot \left( \frac{P_{\text{calon}}}{P_{\text{calon}} - P_{\text{caloff}}} T_{\text{cal}} - T_{\text{cal}} \right) + w_2 \cdot \frac{P_{\text{caloff}}}{P_{\text{calon}} - P_{\text{caloff}}} T_{\text{cal}},
\end{equation}
where the weights \( w_1 \) and \( w_2 \) were determined by the noise variances (\( \sigma_1^2 \) and \( \sigma_2^2 \)). The flux density 
was then calculated as 
\begin{equation}
S = \frac{T_A}{G}, 
\label{4}
\end{equation}
where \( G \) denotes the gain, as provided by \citet{2023RAA....23g5016Z}.
Baseline subtraction was performed by fitting the spectral data with a 
low-order polynomial. 
 



\section{RESULTS} 

We aim to detect spectral lines from quasi-symmetric PAHs
with frequencies defined by $\nu_J = \left(J + \frac{1}{2}\right)\Delta \nu_{\text{comb}}$. For two comb-like spectra, they become indistinguishable if their tooth spacings are excessively close to each other relative to the spectral resolution ($\Delta \nu_{\text{res}}$). The combs are distinguishable when
\begin{equation}
    \Delta \nu_J = \nu_J \cdot \Delta\log(\Delta\nu_{\text{comb}}) \geq \Delta \nu_{\text{res}},
    \label{eq:5}
\end{equation}
i.e., if
\begin{equation}
    \Delta\log(\Delta\nu_{\text{comb}}) \geq \frac{\Delta \nu_{\text{res}}}{\nu} \approx 1 \times 10^{-4}.
    \label{eq:quadratic}
\end{equation}
To ensure no comb lines within the resolvable spectral range are missed, the frequency spacing between adjacent comb lines in the search is set to be smaller than the minimum resolvable comb spacing frequency difference ${\Delta \nu_{\text{res}}}/{\nu}$. We therefore adopt a step size of $\Delta\log(\Delta\nu_{\text{comb}}) = 10^{-5}$, guaranteeing no comb signal is overlooked within the parameter space.

We focused on retrieving PAHs with rotational spectral line peaks in the 1.6–3.2 GHz range. At lower frequency ranges, the beam-dilution effect is more severe, and PAH emissions are presumably weaker. Using a hypothesized rotational state probability distribution, we calculated and then obtained the SNRs for spectra with comb spacing parameters $\Delta \nu_{\text{comb}} = 5$–$10$ MHz. According to Equation~\ref{eq:1}, this parameter range corresponds to PAH molecules containing approximately 138-–194 carbon atoms.

\begin{figure*}
     \centering
     \includegraphics[width=0.8\linewidth]{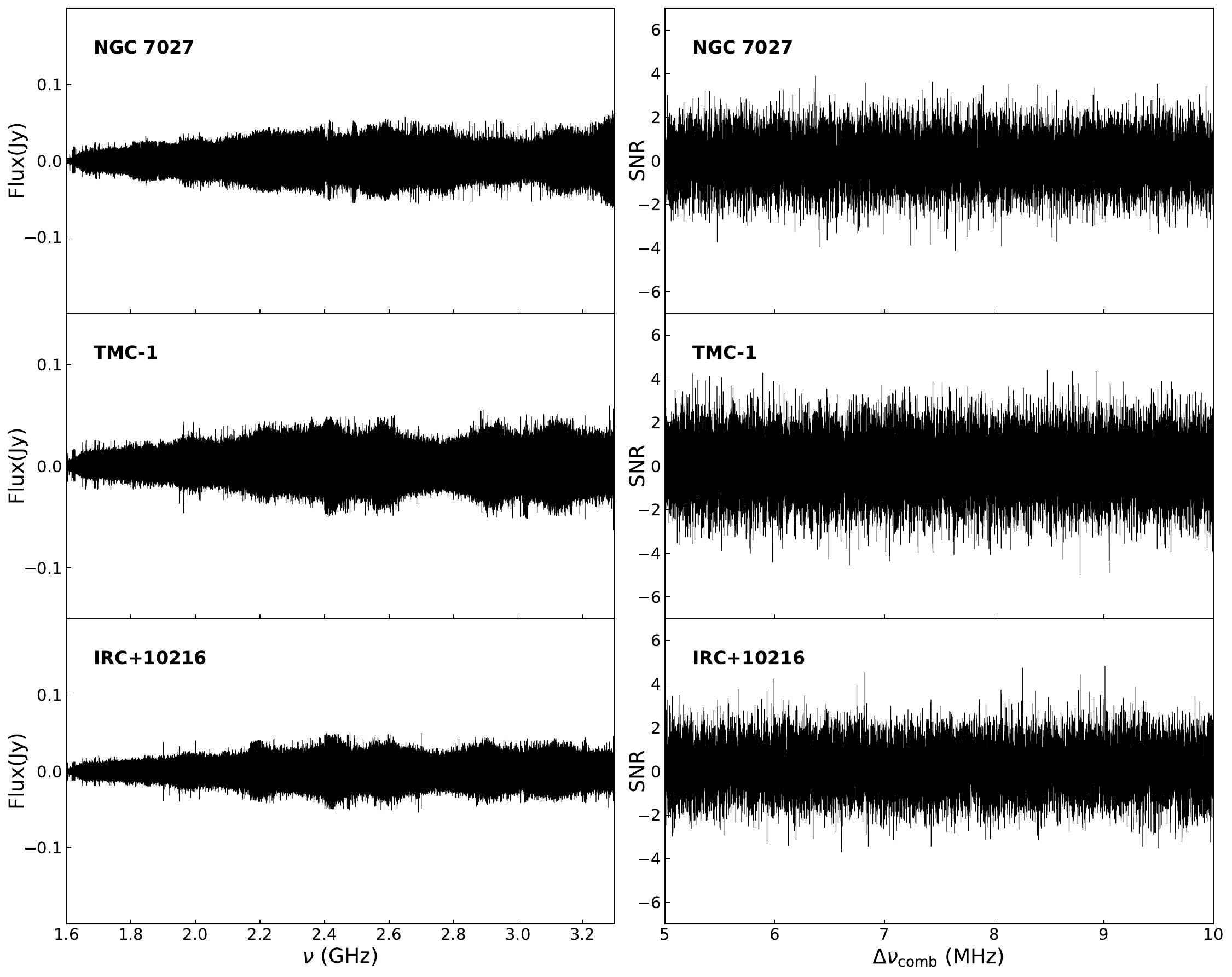}
     \caption{Left: FAST spectral data; Right: SNRs as functions of $\Delta\nu_{\text{comb}}$ derived from our analysis.}
     \label{fig:High_Res_Spectrum}
 \end{figure*}

The search results for the three sources are presented in Fig.~\ref{fig:High_Res_Spectrum}. Notably, no cases where the SNR exceeded the threshold of 5 were identified across the parameter space, with the maximum observed SNR reaching $4.1 \pm 0.3$. There is no significant difference between the spectral results for NGC 7027, TMC-1, and the non-PAH reference source IRC+10216. Results of the PAH spectral detection analysis are statistically indistinguishable from pure noise. Therefore, we did not detect rotational spectra of large quasi-symmetric 
PAHs under current observational sensitivity.

\section{DISCUSSION}

 While no large PAHs were detected in this study, the spectroscopic data and methodology employed enable the derivation of stringent abundance upper limits. Conventionally, abundance upper limits are derived through comparisons between simulated and observational spectra. To predict the expected spectra within the observed bandwidth, one needs to model the rotational emission spectra of PAHs. For this purpose, it is essential to establish the rotational populations of the targeted PAHs. Previous studies have achieved significant progress in formulating this computational framework by extending classical rotational excitation theories \citep{1998ApJ...508..157D}. Subsequent studies \citep{2009MNRAS.395.1055A,2010A&A...509A..12Y,2011ApJ...741...87H,2011MNRAS.411.2750S} have systematically constructed a sophisticated theoretical framework for characterizing rotational state distributions. Under ideal conditions, this framework enables quantitative predictions of PAH rotational spectra through precise characterization of three key components, which include parameters describing the local environment, molecular dipole moments, and structural asymmetry coefficients. Such predictions could subsequently constrain upper limits for PAH column densities~\citep{2025MNRAS.538.2084M}. However, since this search represents a blind exploration focused on large PAHs with diverse carbon numbers, substantial uncertainties persist across multiple parameters. As a result, we deliberately refrain from adopting this methodology.

Instead, we choose to reproduce the rotational spectra of PAHs by implementing the methodological framework put forward by \citet{2014MNRAS.437.2728A} and \citet{2015MNRAS.447..315A}. 
The approach is founded on the assumption that collective rotational emission from typical PAHs, which behave analogously to spinning dust, makes a substantial contribution to the continuum emission. Additionally, it assumes that the rotational states and dipole moments of quasi-symmetric PAHs are analogous to those of typical PAHs.
Although these assumptions are oversimplified, they provide computational tractability.
As noted in \citet{2015MNRAS.447..315A}, the dipole moments of 
distinct nitrogen-substituted PAHs can exhibit variations of up to a factor of 7. This discrepancy may introduce uncertainties as high as 50-fold in estimates of their abundance. According to \citet{2015MNRAS.447..315A}, the flux density of
the spectral line from quasi-symmetric PAHs is described by
\begin{equation}
    S_{\nu}^{\text{PAH}} \approx \frac{N_{\text{PAH}}}{N_{\text{PAH,tot}}} \cdot \frac{\Delta \nu_{\text{comb}}}{\Delta \nu_{\text{res}}} S_{\nu}^{\text{rot}}, \label{eqflux}
\end{equation}
where  \( S_{\nu}^{\text{rot}} \) denotes the flux density of the continuum emission from spinning typical PAHs, 
\( N_{\text{PAH}} \) represents the column density of the specific quasi-symmetric PAH,
and \( N_{\text{PAH, tot}} \) refers to the total column density of all typical PAHs.

 It has been established that the continuum flux densities of NGC 7027 and TMC-1 are dominated 
by free-free radiation \citep{2008ApJ...676..390B,2019MNRAS.486..462P}.
To obtain \( S_{\nu}^{\text{rot}} \), we further
assume that the residual fluxes, after subtracting free-free radiation, are entirely attributed to the emission from spinning typical PAHs.
 For NGC 7027, the flux density of free-free emission (\( S_{\nu}^{\text{ff}} \)) is estimated under the optically thick assumption through the formula:
\begin{equation}
    S_{\nu}^{\text{ff}} = \frac{2k T_{\text{b}} \Omega \nu^2}{c^2}
\end{equation}
where \( k \) is the Boltzmann constant, \( c \) is the speed of light, the brightness temperature \( T_{\text{b}} \approx 11000 \, \text{K} \) \citep{2008ApJ...676..390B}, and the solid angle of the free-free emission region \( \Omega \approx 2.0 \times 10^{-9} \, \text{sr} \) \citep{2003MNRAS.340..381B}.
As a result, the free-free emission flux density of NGC 7027 at 1.7 GHz is 1.78 Jy.  
  The total flux density  of NGC 7027 at 1.7 GHz, as reported in~\citet{2013ApJS..204...19P}, is approximately 2208 mJy with an uncertainty of 8.6 mJy.  Taking into account the beam-dilution effect, with the angular diameter of the molecular emission region of NGC 7027 being \(70''\)
 \citep{1991ApJ...379..271B}
 and the FAST beam size at 1.7 GHz being \(2.37'\), we derived \(S_{\nu}^{\text{rot}} = 83.5 \, \text{mJy}\).
TMC-1 is an extended molecular cloud with an angular size much larger than the FAST beam. \citet{2019MNRAS.486..462P} performed a decomposition of the spectral energy distribution of TMC-1. The uncertainties from the spectral fitting 
represent the upper limit of  \( S_{\nu}^{\text{rot}}\). We derived
\( S_{\nu}^{\text{rot}} \leq 5 \, \text{Jy} \).

\begin{figure*}
        \includegraphics[width=1.8\columnwidth]{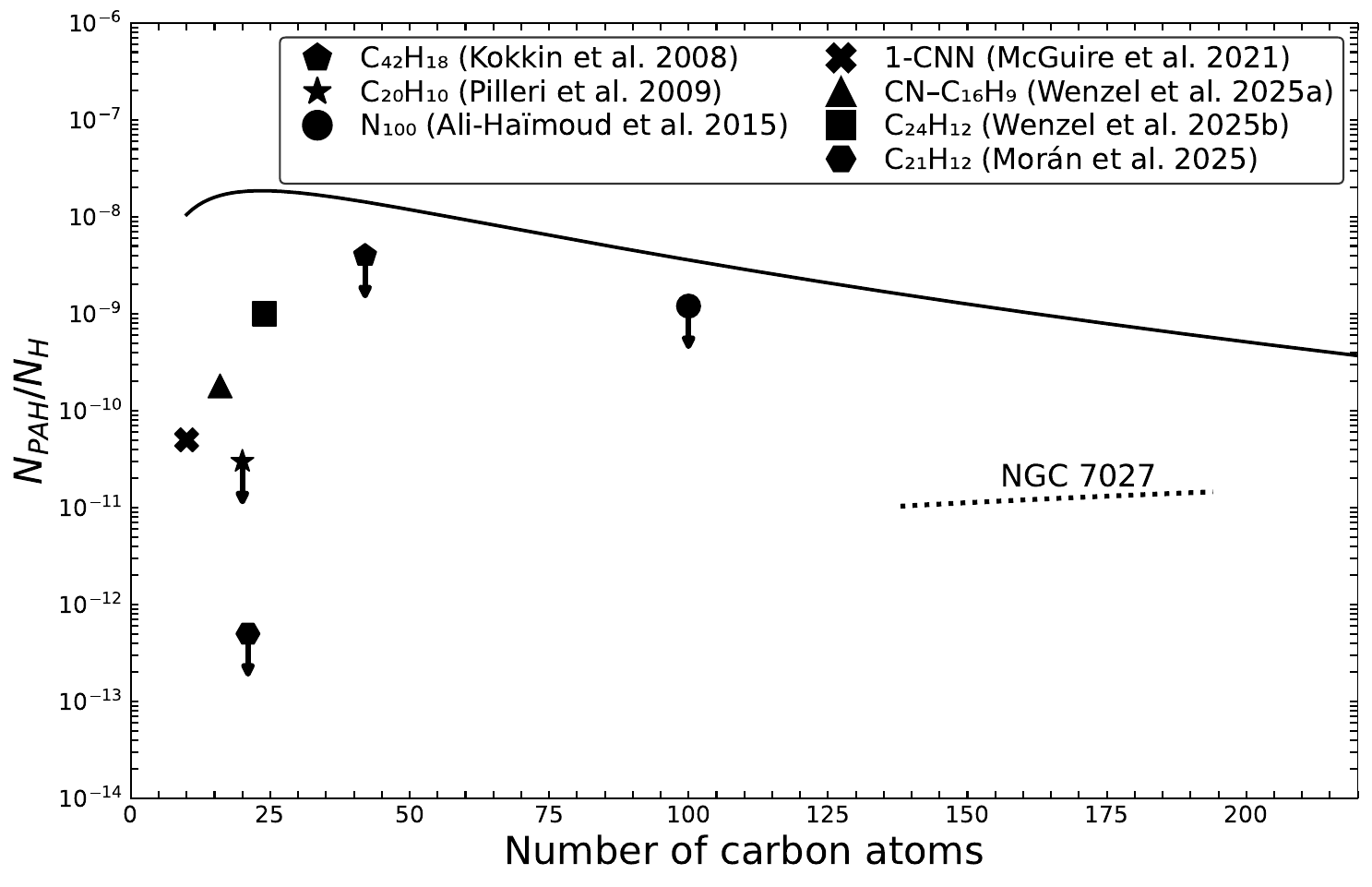}
     \caption{Fractional abundances of PAHs  as a function of the number of carbon atoms $N_{\rm C}$. 
Dotted lines denote the estimated abundance upper limits of large quasi-symmetric PAHs in NGC 7027, and these values warrant cautious interpretation (refer to the contextual discussion). The solid curve shows the theoretical prediction from \citet{1998ApJ...508..157D}. For comparative analysis, values or upper limits from previous studies are overplotted.}
 
    \label{fig:PAH_distribution_color_with_C21H12_2col}
\end{figure*}

Given the achieved sensitivity of $\sigma = 1.6$ mJy, we are able to establish upper limits on the abundance of 
quasi-symmetric PAHs . No SNR exceeding 5 was detected in the comb-like spectra search, which implies that:
\begin{equation}
    \frac{S_{\nu}^{\text{PAH}}}{1.6 \, \text{mJy}} \sqrt{\frac{\Delta \nu_{\text{tot}}}{\Delta \nu_{\text{comb}}}} \leq 5, \label{upper_bound}
\end{equation}
where $\Delta \nu_{\text{tot}}$ denotes the total frequency range covered by our observations, and thus 
${\Delta \nu_{\text{tot}}}/{\Delta \nu_{\text{comb}}}$ denotes the number of comb-like spectral components. By combining Equations~(\ref{eqflux}) and (\ref{upper_bound}), we derive:
\begin{equation}
    \frac{N_{\text{PAH}}}{N_{\text{PAH,tot}}} \lesssim 0.225 \times \sqrt{\frac{1 \, \text{Hz}}{\Delta \nu_{\text{comb}}}} \times \frac{1 \, \text{mJy}}{S_{\nu}^{\text{rot}}}.
\end{equation}
Given that our search range $\Delta \nu_{\text{comb}}$ spans from 5 MHz to 10 MHz, the upper limits on the $N_{\text{PAH}}/N_{\text{PAH,tot}}$ ratios are found to range from $8.5  \times 10^{-7}$ to $1.2 \times 10^{-6}$ for NGC 7027. 
Assuming 20\% of total carbon atoms are sequestered in PAHs, and thus the fraction of carbon locked in PAHs relative to hydrogen is \( 6 \times 10^{-5} \)~\citep{1992ApJ...393L..79J,2001ApJ...554..778L}, we estimate the abundances of quasi-symmetric PAH  relative to hydrogen as \( {N_{\text{PAH}}}/{N_{\rm H}} 
\lesssim 1{-}1.4 \times 10^{-11} \), as shown in Figure~\ref{fig:PAH_distribution_color_with_C21H12_2col}. The upper limit of \( N_{\rm PAH}/N_{\rm H} \) should be interpreted with caution, as radiations other than free-free emission and spinning PAHs may contribute to the continuum emission. If \( S_{\nu}^{\text{rot}} \) is lower than 83.5\,mJy, the upper limit of \( N_{\rm PAH}/N_{\rm H} \) would increase accordingly. For TMC-1, a meaningful upper limit of \( N_{\rm PAH}/N_{\rm H} \) cannot be derived, given that only a loose upper limit of \( S_{\nu}^{\text{rot}} \) has been estimated.

By superimposing an additional independent component that characterizes carbonaceous grains (predominantly PAHs) onto the classical MRN power-law distribution \citep{1977ApJ...217..425M}, \citet{1998ApJ...508..157D} derived a theoretical PAH abundance as a function of PAH radius \( a \). This supplementary component follows a lognormal distribution, which aligns naturally with physical intuition for grain populations governed by stochastic growth or fragmentation processes. Given that the formation mechanisms of PAHs and their evolutionary pathways typically exhibit inherent stochasticity, grain sizes manifest a normal distribution in logarithmic space; a similar distribution has also been presented by \citet{1990A&A...237..215D}.
 The theoretical prediction of \citet{1998ApJ...508..157D} is overplotted in Figure~\ref{fig:PAH_distribution_color_with_C21H12_2col} through the conversion relation \( a = 0.9 \sqrt{N_{\rm C}} \, \text{\AA} \), which holds for fused-ring PAHs \citep{1986A&A...164..159O}. In this figure, we further compare the abundance or abundance upper limits of smaller PAHs derived from the literature (see Appendix~\ref{appa} for detailed descriptions) with theoretical predictions. It appears that both small and large PAHs have abundances lower than the theoretical expectations.

Fully attributing the UIE bands to PAHs yields an estimated PAH abundance relative to \({\rm H}_2\) of \(3 \times 10^{-7}\) \citep{2008ARA&A..46..289T}, a value notably higher than those derived from PAH rotational spectra. \citet{2009MNRAS.397.1053P} attributed the depleted abundance of C\(_{20}\)H\(_{10}\) in a circumstellar
envelope to the conversion of small PAHs into larger ones. However, our analysis indicates that large PAHs also appear to be depleted. This may cast some doubt on the hypothesis that PAHs are the carriers of UIE bands, and instead lends plausible support to MAON-like materials as the UIE carrier. 
Such materials have an amorphous three-dimensional structure, the inherent characteristics of which result in their rotational spectra being distinctly different from the comb-like spectra targeted in this study.

\section{CONCLUSION}
In astronomical literature, UIE  has been commonly referred to as a synonym for PAH emission. However, it is crucial to emphasize that the assignment of UIE carriers to PAHs remains far from conclusive. 
 Hopefully, conclusive evidence for this assignment will emerge from the discovery of rotational spectral lines of large PAHs, even though small PAHs have already been definitively detected. Here, we present the most sensitive spectra in the 1.6–3.2 GHz range toward two UIE sources (NGC 7027 and TMC-1) and a non-UIE source (IRC +10216), aiming to search for comb-like spectral signatures from quasi-symmetric PAHs. The observed frequency range enables us to probe emissions from large PAHs containing 138–194 carbon atoms. To this end, we employed a matched filter technique 
which can identify weak periodic signals amidst noisy spectra. Despite the highly sensitive observations and dedicated spectral reduction methods, no comb-like features with SNR$>5\sigma$ were detected.

Do the FAST spectroscopic observations reveal a scarcity of large PAHs in astronomical environments?
The conclusion remains inconclusive.
Although the upper limit on the abundances of large PAHs derived from these observations appears lower than theoretical expectations and those inferred from UIE  observations, the abundance constraints require extreme caution in interpretation. Their validity critically depends on two assumptions: first, that a subset of PAHs exhibits perfectly harmonic spectral signatures; second, that typical PAHs, if present, should collectively contribute to continuum emission. Neither of these assumptions has been empirically validated. Nevertheless, this study offers valuable observational constraints for exploring alternative UIE carriers, such as MAON-like materials.

\section*{Acknowledgements}
We gratefully acknowledge the anonymous referee for his/her valuable comments, which have significantly enhanced the quality of this paper.
The financial supports of this work are from 
the National Natural Science Foundation of China (NSFC, No.\,12473027 and 12333005), the Guangdong Basic and Applied Basic Research Funding (No.\,2024A1515010798), and the Greater Bay Area Branch of the National Astronomical Data Center (No.\,2024B1212080003). 
This article is based
upon work from COST Action CA21126 - Carbon molecular nanostructures in space (NanoSpace), supported by COST (European Cooperation in Science
and Technology).
This work made use of the data from FAST (Five-hundred-meter Aperture Spherical radio Telescope). FAST is a Chinese national mega-science facility, operated by National Astronomical Observatories, Chinese Academy of Sciences.

\section*{Data Availability}

The data underlying this article will be shared on reasonable request to the corresponding author.



\bibliographystyle{mnras}
\bibliography{example} 




\appendix
\section{PAH abundances from previous studies}
\label{appa}

This appendix compiles PAH abundances from previous studies. \citet{2008ApJ...681L..49K} and \citet{2009MNRAS.397.1053P} estimated abundance upper limits for $\text{C}_{42}\text{H}_{18}$ and $\text{C}_{20}\text{H}_{10}$ relative to hydrogen in the ISM and the Red Rectangle nebula, respectively. For TMC-1, observational analyses have yielded column densities or upper limits for four molecular species: $\text{C}_{21}\text{H}_{12}$, CN--$\text{C}_{16}\text{H}_{9}$, $\text{C}_{24}\text{H}_{12}$, and 1--cyanonaphthalene (1--CNN)~\citep{2025MNRAS.538.2084M,2025NatAs...9..262W,2025ApJ...984L..36W,2021Sci...371.1265M}. These abundances (or upper limits) were determined using an established H$_{2}$ column density of $N(\text{H}_{2}) = 1 \times 10^{22}\ \text{cm}^{-2}$~\citep{2016ApJS..225...25G,2019A&A...624A.105F}. \citet{2015MNRAS.447..315A} derived an $N_{\rm PAH}/N_{\rm PAH,tot}$ upper limit for quasi-symmetric PAHs with $\sim$100 carbon atoms in TMC-1, using the same methodology adopted in this study. This value was scaled to obtain abundance upper limits. These abundance and abundance upper limits of PAHs are incorporated in Figure~\ref{fig:PAH_distribution_color_with_C21H12_2col}.



\bsp	
\label{lastpage}
\end{document}